\documentclass[preprint,graphicx]{revtex4-1}

\usepackage{graphicx}

\draft

\begin{document}
	
	\title{Tuning Strain in Flexible Graphene Nanoelectromechanical Resonators}
	\author{Fen Guan}
	\author{Piranavan Kumaravadivel}
	\author{Dmitri V. Averin}
	\author{Xu Du}
	%\Email{xu.du@stonybrook.edu}
	\affiliation{Department of Physics \& Astronomy, Stony Brook University, Stony Brook, NY, 11790, USA	}

    \date{\today}
	
\begin{abstract}
The structural flexibility of low dimensional nanomaterials offers unique opportunities for studying the impact of strain on their physical properties and for developing innovative devices utilizing strain engineering. A key towards such goals is a device platform which allows the independent tuning and reliable calibration of the strain. Here we report the fabrication and characterization of graphene nanoelectromechanical resonators(GNEMRs) on flexible substrates. Combining substrate bending and electrostatic gating, we achieve the independent tuning of the strain and sagging in graphene and explore the nonlinear dynamics over a wide parameter space. Analytical and numerical studies of a continuum mechanics model, including the competing higher order nonlinear terms, reveal a comprehensive nonlinear dynamics phase diagram, which quantitatively explains the complex behaviors of GNEMRs.
\end{abstract}

\maketitle

Understanding and controlling the effect of strain on the physical properties of low dimensional nanomaterials has been a topic of great interest in recent years. Various types of strain have been theoretically predicted to exert fundamental impact on the electronic properties, including flexural phonon scattering\cite{Castro2010a}, gauge field formation\cite{Pereira2009} and modification of lattice symmetry\cite{Pereira2009a}. Through strain engineering, the mechanical, electronic, and optical properties of a nanomaterial may be tailored to better suit its functionality. While extensive work has been carried out to study those properties via optical\cite{Ni2008}\cite{Yoon2011} and local probe techniques\cite{Levy2010}\cite{Lopez-polin2015}, it is ultimately desirable to achieve strain engineering in electronic devices\cite{Petrone2013}. The major hurdles in achieving this goal are the absence of an independent tuning knob for strain in an electronic device and a reliable approach for characterizing the strain. 

A promising device platform for realizing strain engineering is a suspended field effect transistor (FET)-like nanoelectromechanical resonator(NEMR). Graphene, for example, has been intensively studied in NEMR devices\cite{Bunch2007}\cite{Chen2013a}, its large mechanical strength\cite{Lee2008}, high elastic modulus\cite{Frank2007} and low mass density. With graphene suspended from the substrate, these devices potentially allow probing its intrinsic properties. In addition, the strain in graphene can be tuned electrostatically through the gate bias. And the magnitude of the strain can be estimated from the gate dependence of the resonant frequency of graphene.\cite{Chen2009}. 

Despite these advantages, a major limitation of tuning strain through electrostatic force is that it unavoidably affects charge doping and hence the electronic properties of graphene. Moreover, the electrostatic strain is always tensile. The range of such strain tuning is also limited(\textasciitilde0.01\% at 10V\cite{Bao2012a}), because large gate bias can cause collapse of graphene channel. This tuning range is usually much smaller than the residual strain from the fabrication process (\textasciitilde 0.1\%\cite{Metten2013}). 

Electrostatic tuning can also complicate the analysis of strain by inducing strong nonlinearity under large gate bias. Typically, large deformation\cite{Yakobson1996}, clamping geometry\cite{Lifshitz2008} and external force\cite{Eichler2011a} introduce higher order terms of displacement and restoring force. These terms play significant roles in the equation of motion and give rise to deviation from simple harmonic behavior. In FET-like GNEMRs, tuning the strain by perpendicular electrostatic force inevitably causes the sagging of graphene, which affects the resonance dynamics. Under large driving amplitude, broadened nonlinear resonance has been observed. Depending on the random residual strain and sagging, the broadened resonance shifts either to higher frequencies (i.e. hardening behavior)\cite{Chen2009} or to lower frequencies (softening behavior)\cite{Song2012} with increasing oscillation amplitude. A comprehensive study of the relation between the strain and nonlinear dynamics is therefore important for determining strain in the nonlinear regime. However, further experimental study has been restricted, due to the lack of an effective tuning knob for the strain\cite{Kozinsky2006}.

Here we develop a device platform for strain engineering, by fabricating FET-like local-gated GNEMRs on flexible substrates. Bending the substrate enables the independent tuning of the strain. By analyzing the gate bias and substrate bending dependence of the nonlinear dynamics, the strain in graphene is reliably determined. 

\begin{figure}
	%\centering
	\includegraphics{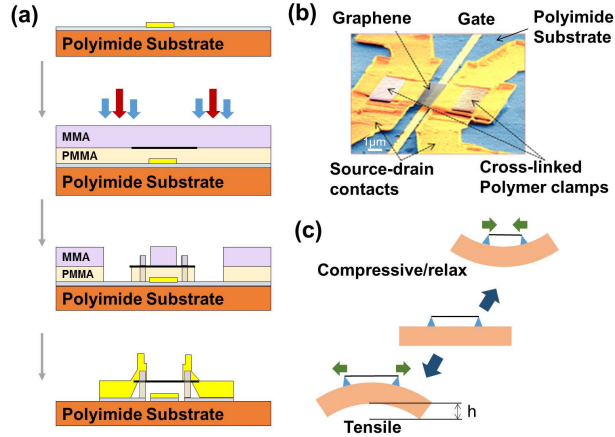}
	\caption{
		\textbf{Device fabrication and strain tuning.}
		\textbf{(a)} Process flow diagram for device fabrication.	\textbf{(b)} False color scanning electron micrope (SEM) image of a suspended graphene sheet with Au contacts and bottom gate. \textbf{(c)} Mechanism for strain tuning by bending the substrate. $h$ is the deflection of substrate, which is negative for compressive surface strain, positive for tensile strain.}
	\label{fig:fabrication}
\end{figure}

The flexible substrate used here is 125 $\mu$m thick Polyimide, which maintains flexibility over a wide temperature range (\textasciitilde 4K - 650K). The fabrication of a GNEMR on the Polyimide substrate is depicted in Fig.~\ref{fig:fabrication}(a). First, we evaporate 6 nm of Ti onto the Polyimide substrate to avoid the static charging in electron beam lithography (EBL). After patterning the substrate with 1$\mu$m wide, 30 nm thick gold gate electrodes by EBL, we coat the substrate with 300 nm polymer spacer (PMMA). Simultaneously, a polydimethylsiloxane/transparent tape/methylmethacrylate(MMA) stack is prepared on a transparent glass slide\cite{Hunt2013}, with graphene flakes exfoliated onto MMA. Using a micro-manipulator, the MMA membrane with graphene flakes is transferred onto the PMMA covered substrate, with the chosen graphene flake aligned to the gate electrode. Then following an etching-free suspending procedure\cite{Mizuno2013}, we fabricate the electrical contact pads using EBL. Part of the polymer layers (both MMA and PMMA) are cross-linked so that the graphene is strongly clamped at the contacts\cite{Min2014}\cite{Zailer1999}. Following metalization with Cr(1 nm)/Au(200 nm), the standard ``lift-off" process is carried out using acetone, after which the graphene is supported by the 3D metal contact ``pillboxes". In the end the exposed thin Ti conducting layer is etched away using diluted HF acid. Throughout the whole lift-off and etching process, the device goes through the following solvents successively: acetone, isopropyl alcohol(IPA), de-ionized(DI) water, diluted HF, DI water and IPA. Finally, it's taken directly out of hot Hexane. From the thickness of the polymer spacer and the gate electrode, the distance between graphene and the gate $d$ is estimated to be 270 nm. Fig.~\ref{fig:fabrication}(b) shows the SEM image of a finished device where graphene is suspended over the substrate by the 3D structure of the metal contacts, clamped by cross-linked polymer. In all our devices, we select rectangular graphene flakes with sharp and straight edges to avoid edge modes and local buckling\cite{Garcia-Sanchez2008}. Bending the substrate with a deflection $h$ (positive for outward bending and negative for inward bending), the surface strain of the substrate modifies the source-drain distance, hence tuning strain in graphene (Fig.~\ref{fig:fabrication}(c)).

For resonance measurements, we have adopted the electrical mixing method reported in a previous work\cite{Chen2009}. In this method, the oscillation is characterized by the mixed current through graphene (excluding background), which is proportional to the oscillation amplitude of the graphene membrane. The measurement is carried out in vacuum below 50 mTorr at room temperature. The deflection of the substrate $h$ is controlled by adjusting a plunger with a threaded rod, which is coupled to the ambient through a vacuum rotary feed-through. The detailed setup and the circuit diagram are shown in the Supplementary Material\cite{Supplementary}. All measurement results presented in this paper are obtained from a device with graphene channel of width $W= 2\ \mu$m and length $L=1\ \mu$m.

\begin{figure}
	%\centering
	\includegraphics{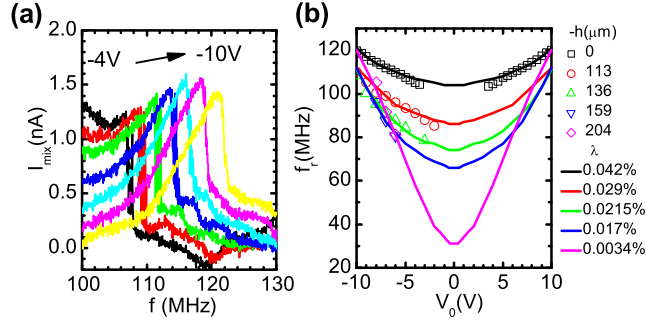}
	\caption{\textbf{Dependence of resonant frequency on gate bias.} \textbf{(a)} Mixed current (proportional to the oscillation amplitude) versus frequency at various DC gate bias V$_0$. As the bias voltage changes from -4V (black) to -10V (yellow), the resonant frequency increases monotonically. \textbf{(b)} Experiment data and simulation results of resonant frequency versus gate bias at various values of $\lambda$.}
	\label{fig:vsVg}	
\end{figure}

We first study the resonance of the GNEMR device without bending the substrate. Fig.~\ref{fig:vsVg}(a) shows the mixed current as a function of frequency with gate bias $V_0$ tuned from -4 V to -10 V. The corresponding resonant frequency $f_r$ increases monotonically from 108 MHz to 122 MHz. Sensitivity of $f_r$ on $V_0$ indicates that the observed resonance is from the graphene and not from the electrodes (which is almost $V_0$ - independent). Both the large values of $f_r$ and the lineshape of hardening behavior imply significant tensile residual strain in graphene. The resonant frequencies for various gate bias values are shown in Fig.~\ref{fig:vsVg}(b) (see the data labeled by squares). The values are symmetric around $V_0=0$ and qualitatively consistent with previous report\cite{Chen2009}. For the device studied here, since the hole branch shows higher transconductance and gives better resonance signal than the electron branch (see Fig.S1(b) in the Supplementary Material\cite{Supplementary}), we focus on the measurement at negative gate biases.  
 
\begin{figure}
	%\centering
	\includegraphics{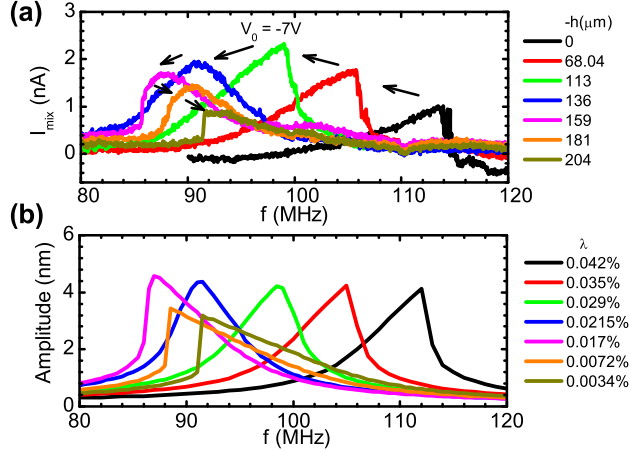}
	\caption{\textbf{Dependence of resonant frequency and lineshape on strain $\lambda$.} Different color corresponds to different $\lambda$. \textbf{(a)} Experiment results. The direction of the arrows indicates increasing inward bending (i.e larger $|h|$ and smaller $\lambda$). \textbf{(b)} Simulation results. The simulated frequency and lineshape of resonance peaks match with the experiment results. Both experiment and simulation results show non-monotonic dependency of resonant frequency and transition from hardening to softening resonance when $\lambda$ is decreased close to zero.}
	\label{fig:vsbend} 
\end{figure}

\begin{figure}
	%\centering	
	\includegraphics{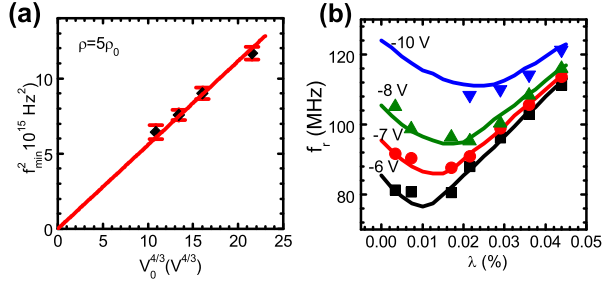}
	\caption{\textbf{Dependence of resonant frequency on gate bias and $\lambda$}. Solid lines are fits from model. Symbols are experiment data. \textbf{(a)} Linear fitting of $f_{min}^2$ versus $V_0^{4/3}$. \textbf{(b)} Resonant frequency $f_r$ versus strain $\lambda$ at various gate biases.}
	\label{fig:fvs}	
\end{figure}

Next we focus on how the resonance depends on the strain $\lambda$. Here we define $\lambda$ as the strain in graphene at zero gate bias, which takes into account both the residual tension and the substrate bending. The tensile residual strain indicates $\lambda >0$ without bending. Fig.~\ref{fig:vsbend}(a) shows the mixed current as a function of the frequency at $V_0=-7V$, as the tensile strain  is reduced by inward bending of the substrate. Here the resonant frequency decreases from 114 MHz to 86 MHz as $|h|$ increases to 159 $\mu m$. Further inward bending results in the increasing of $f_r$ up to 92 MHz. This non-monotonic dependence of $f_r$ on $\lambda$ generally holds for different $V_0$, as summarized in Fig.~\ref{fig:fvs}(b). Moreover, as $V_0$ is lowered, $\lambda$ corresponding to the minimum resonant frequency gets smaller. Here $\lambda$ is quantitatively calibrated through numerical simulation discussed later. We note that the observed $\lambda$ and $V_0$ dependence of the resonant frequency is reproducible upon repeated inward bending and releasing of the substrate, indicating a good mechanical clamping of the graphene channel and an absence of structural slippage.

In conjunction with the non-monotonic behavior of the resonant frequency, the resonance lineshape also undergoes a transition around the minimum resonant frequencies in Fig.~\ref{fig:vsbend}(a). Without substrate bending, the resonance shows a hardening behavior. As $|h|$ increases to 136 $\mu m$, the resonance peak becomes symmetric. At the minimum $f_r$, the softening resonance emerges. The softening behavior gets increasingly pronounced with further inward bending, as indicated by the steeper rise of the resonance amplitude at the tipping point. While softening and hardening behaviors have been randomly observed in GNEMRs\cite{Chen2009}\cite{Song2012}, here we achieve fine control of the nonlinear dynamics by combining lateral stretching and vertical electrostatic pulling.

\begin{figure} 
	%\centering
	\includegraphics{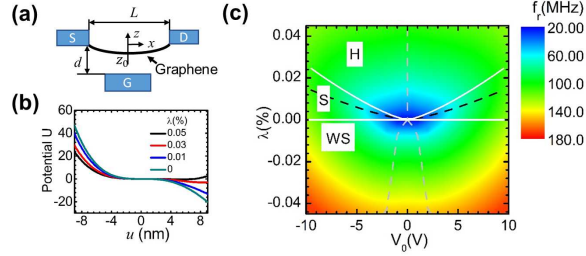}
	\caption{	\textbf{Continuum mechanics model for doubly clamped graphene resonator.} \textbf{(a)} Schematic of the graphene resonator. $z_0$ is the static displacement of the center of graphene. $L$ is the distance between S-D electrode clamps. The shape of graphene is described by z(x). $d$ is the distance between S-D and gate. \textbf{(b)} Nonlinear contribution to the potential energy plotted as a function of average oscillatory displacement $u$. \textbf{(c)} Nonlinear dynamics phase diagram. Three characteristic regimes - hardening (H), softening(S) and the non-monotonic softening (WS) regime - are separated by the white solid lines on the diagram. The black dashed line highlights the frequency minima at fixed gate biases and is close to the hardening-to-softening transition line. The grey dashed lines correspond to the frequency minima at fixed $\lambda$.}
	\label{fig:Model}
\end{figure}

To quantitatively understand the observed resonance behavior, we model graphene as a continuum membrane with 1D approximation\cite{Sapmaz2003}\cite{Atalaya2008}\cite{Chen2013}\cite{Westra2010}, which is illustrated in Fig.~\ref{fig:Model}(a). We consider the driving gate voltage to be $V(t)=V_0+\delta Vcos(\omega t)$, under which the deflection of graphene is approximately a combination of a parabolic static deflection and an oscillating fundamental mode: $z(x,t)=z_0(1-4x^2/L^2)+u(t)cos(\pi x/L)$. Here $z_0$ is the displacement at the center of graphene, $L$ is the source-drain distance. Assuming the strain-free length of the graphene beam is $l_0$, the strain $\lambda \equiv (L-l_0)/l_0$. We then obtain the following nonlinear differential equation for the oscillatory component $u$:
\begin{equation}
\rho \frac{\partial^2u}{\partial t^2} + \eta \frac{\partial u}{\partial t} +ku+Au^2+Bu^3+ \frac{4\varepsilon_0 V_0 \delta Vcos(\omega t)}{\pi d^2}=0 \,.
\label{equ} \end{equation}
Here $\rho$ is the 2D mass density, $\eta$ characterizes the damping strength of the resonator, $d ( \gg z)$ is the vertical distance between the source-drain and the gate electrode, $k=\pi^2 EL^{-2}[\lambda+(8/3+{512}/\pi^4)z_0^2L^{-2}]$, $A=16\pi E z_0L^{-4}$ and $B=(\pi^4/4)EL^{-4}$, in which $E$ is the 2D Young's modulus of graphene. Note that these expressions assume, as appropriate for the geometry of our device, that $z_0$ and $u$ in the graphene deflection are much smaller than the distance $d$ to the gate electrode. In this regime, electrostatic energy of the graphene-gate capacitance gives only negligible contribution to the coefficients $k$, $A$, $B$. In particular, for the effective spring constant $k$, the electrostatic contribution: $\varepsilon_0 V_0d^{-3}=16z_0/dEL^{-2}[\lambda+8/3z_0^2L^{-2}]$, is of the order of $z_0/d\ll1$ in relative to the elastic contribution. If $z_0, u \sim d$, electrostatic energy can change the behavior of the resonance\cite{Sillanpaa2011}.  

The main resonant frequency can be characterized by $f_r=1/2\pi\sqrt{k/\rho}$. To qualitatively understand how $f_r$ varies with the strain $\lambda$, we can consider the dependence of the effective stiffness of the graphene channel on the deformation along the z-axis, described by an effective spring constant $k_{eff} \simeq \partial (T\partial^2z/\partial x^2)/\partial z)|_{x=0}$. The presence of a minimum frequency can be understood considering the competition between the tension $T$ and the curvature $\partial^2z/\partial x^2$. When $\lambda>0$ and is large, the sagging is small and the oscillatory motion is perpendicular to the direction of tension. The behavior of the resonator is similar to that of a classical string whose resonant frequency decreases with decreasing tension. On the other hand, when $\lambda$ gets smaller, the sagging gets larger, increasing the curvature of graphene. This significantly couples the oscillatory motion with the longitudinal deformation. As a result, $f_r$ starts to increase with decreasing $\lambda$. A more rigorous analysis taking into account the oscillatory component (see the Supplementary Material\cite{Supplementary}) finds the minimum frequency and the corresponding $\lambda$ to be:
\begin{eqnarray}
\label{eqfmin}
 f_{min}=\frac{1}{2\pi}\big(\frac{27 \pi^2 \varepsilon_0^2E }{ 2\rho^3 L^4 d^4} V_0^4\big)^{1/6}, \\
\nonumber
\lambda_{min}= \Big(\frac{2}{\pi}-\frac{\pi^3}{192}\Big) \big(\frac{2}{\pi} \big)^{1/3} \big(\frac{\varepsilon_0 L V_0^2}{2Ed^2}\big)^{2/3}.
\end{eqnarray}

The transition from hardening to softening behavior with decreasing $\lambda$ can be understood qualitatively considering the competition between the quadratic and the cubic term in Eq.(\ref{equ}). Under large tension, $z_0 \simeq 0$ and the quadratic term is negligible. So the major contribution to the ``mean" $k_{eff}$ is from the positive cubic term, which increases with increasing amplitude. This gives rise to the hardening Duffing oscillator behavior. On the other hand, under small tension, the presence of a gate bias causes significant sagging, hence the quadratic term dominates. This term induces a tilt in the potential energy, as illustrated in Fig.~\ref{fig:Model}(b). As a result, with increasing oscillatory amplitude, the average oscillatory displacement $u$ shifts towards more positive values, where the curvature of the potential well (the $k_{eff}$) is smaller. This gives rise to a decreasing resonant frequency with increasing amplitude, hence softening behavior. Following a more detailed calculation shown in the Supplementary Material\cite{Supplementary}\cite{Landau1976}, the switching between the two behaviors, in the limit of small driving force, happens at $\lambda= \Big(\frac{2}{\pi}-\frac{\pi^3}{544}\Big)\big( \frac{17}{3\pi}\big)^{1/3} \big(\frac{\varepsilon_0 L V_0^2}{2Ed^2}\big)^{2/3}$. This shows similar dependence on $V_0$ as Eq.(\ref{eqfmin}), but with a slightly larger prefactor.

To obtain a quantitative comparison between the model and the data, we first evaluate the mass density of the sample. In agreement with Eq.(\ref{eqfmin}), the data of $f_{min}^2$ versus $V_0^{4/3}$ follows a linear dependence which extrapolates to the origin(Fig. \ref{fig:fvs}(a)). The slope, which is proportional to $1/\rho$, provides an estimation to the mass density of the channel. Adopting $E_{3d}=1$ TPa\cite{Lee2008} and using the device geometry $d=270 nm$ and $L=1\mu m$, we obtain the mass density to be ~$5\rho_0$, in which $\rho_0=7.4\times10^{-7}kg/m^2$ is the mass density of monolayer graphene. The large mass density, which has been frequently observed in previous work\cite{Chen2009}\cite{Song2012}, can be attributed to the resist residue and other contaminants. Using the estimated mass and Eq.(\ref{equ}), we are able to numerically simulate the gate bias dependence of the resonant frequency at various $\lambda$. For the simulations, we use a dissipation factor of $\eta=50$ and the experimental driving amplitude $\delta V=300mV$, but vary the value of $\lambda$ as a fitting parameter. The simulation results agree well with the data from experiment, as shown in Fig.~\ref{fig:vsVg}(b). We also compare the simulated line shapes of the resonance peaks with the measurement results. As shown in Fig.~\ref{fig:vsbend}(b), the model reproduces our observation in Fig.~\ref{fig:vsbend}(a) remarkably well. The suspended graphene device analyzed here shows a residual strain of \textasciitilde 0.042\%, which is relaxed down to \textasciitilde 0\% through substrate bending. The small discrepancy between the simulation and the experiment results may be attributed to the errors in the device geometry estimation and the simplification of our model, which neglects the nonlinear damping\cite{Eichler2011a} and the higher order modes. 

A comprehensive parametric dependence of the nonlinear dynamics in the GNEMRs is shown in Fig.~\ref{fig:Model}(c). Based on numerical calculations, the color-coded resonant frequency is plotted as a function of gate bias and $\lambda$. Three characteristic regimes can be identified on the diagram. As observed in our devices, the hardening(H)-to-softening(S) transition happens close to the minimum frequency line (dashed black line). The regime, not covered in this work but has been previously reported~\cite{Song2012}, is the WS regime for negative $\lambda$. Here the gate-dependence of resonant frequency become non-monotonic with a ``W" shape. The nonlinear dynamics phase diagram covers the so-far observed behaviors of GNEMRs and may serve as a useful guide to the analysis of 1D-like mechanical resonators in general. 

In summary, we have demonstrated a FET-like suspended graphene nanoelectromechanical resonator on a flexible substrate, where the strain can be independently tuned by bending the substrate. This design allows for the observation of nonlinear dynamics in GNEMRs, including non-monotonic strain-dependent resonant frequency and the transition from hardening to softening resonance behavior. The device platform developed here allows the independent tuning of strain for measurements on the electronic properties of the channel materials, without electric field doping effects. This vastly broadens the scope of the strain-engineering of 2D nanomaterials.

\begin{acknowledgments}
This work was supported by NSF (Grant DMR 1105202) and AFOSR (Grant FA9550-14-1-0405). Part of this research used resources of the Center for Functional Nanomaterials, which is a U.S. DOE Office of Science User Facility, at Brookhaven National Laboratory under Contract No. DE-SC0012704. We thank FRALOCK Corp. for supplying the Polyimide substrate samples.
\end{acknowledgments}

\bibliography{My}

\end{document}